\documentclass[showpacs,preprintnumbers,superscriptaddress,aps,pre,10pt]{revtex4}
\usepackage{amssymb}
\usepackage{amsmath}
\usepackage{times}
\usepackage{graphicx}
\usepackage{epsfig}

\RequirePackage{color}
\newcommand{\red}[1]{{\color{black}{#1}}}

\definecolor{MyDarkGreen}{rgb}{0.02,0.60,0.06}

\graphicspath{{figures}}

\begin{document}

\def\s{\sigma}
\def\t{\tau}
\def\d{\delta}
\def\be{\begin{equation}}
\def\ee{\end{equation}}
\def\o{\omega}
\def\bea{\begin{eqnarray}}
\def\eea{\end{eqnarray}}
\def\L{{\cal L} }

\title[Finite-size corrections and scaling for the dimer model on the checkerboard lattice]{Finite-size corrections and scaling for the dimer model on the checkerboard lattice}

\date{October 14, 2016}

\author{Nickolay Sh. Izmailian}
\email{izmail@yerphi.am}
\affiliation{Yerevan Physics Institute, Alikhanian Brothers Street 2, 375036 Yerevan, Armenia}

\author{Ming-Chya Wu}
\email{mcwu@ncu.edu.tw}
\affiliation{Research Center for Adaptive Data Analysis, National Central University, Zhongli, Taoyuan 32001, Taiwan}
\affiliation{Institute of Physics, Academia Sinica, Nankang, Taipei 11529, Taiwan}

\author{Chin-Kun Hu}
\email{huck@phys.sinica.edu.tw}
\affiliation{Institute of Physics, Academia Sinica, Nankang, Taipei 11529, Taiwan}
\affiliation{National Center for Theoretical Sciences, National Tsing Hua University, Hsinchu 30013, Taiwan}
\affiliation{Business School, University of Shanghai for Science and Technology, Shanghai 200093, China}

\begin{abstract}
Lattice models are useful for understanding behaviors of interacting complex many-body systems. The lattice dimer model has been proposed to study the adsorption of diatomic molecules on a substrate. Here we analyze the partition function of the dimer model on an $2 M \times 2 N$ checkerboard lattice wrapped on a torus and derive the exact asymptotic expansion of the logarithm of the partition function. We find that the internal energy at the critical point is equal to zero. We also derive the exact finite-size corrections for the free energy, the internal energy, and the specific heat. Using the exact partition function and finite-size corrections for the dimer model on finite checkerboard lattice we obtain finite-size scaling functions for the free energy, the internal energy, and the specific heat of the dimer model. We investigate the properties of the specific heat near the critical point and find that specific-heat pseudocritical point coincides with the critical point of the thermodynamic limit, which means that the specific-heat shift exponent $\lambda$ is equal to $\infty$. We have also considered the limit $N \to \infty$ for which we obtain the expansion of the free energy for the dimer model on the infinitely long cylinder. {\red{From a finite-size analysis we have found that two conformal field theories with the central charges $c = 1$ for the height function description and $c = -2$ for the construction using a mapping of spanning trees can be used to describe the dimer model on the checkerboard lattice.}}
\end{abstract}



\maketitle

\section{Introduction}
\label{introduction}

Lattice models are useful for understanding behaviors of interacting complex many-body systems. For examples, the Ising model \cite{onsager,95jpa3dIsing,96jpaIsing} can be used to understand the critical behavior of gas-liquid systems \cite{review,2012jcp}, the lattice model of interacting self-avoiding walks \cite{Orr47,r1,13eplISAW,16cpc} can be used to understand the collapse and the freezing transitions of the homopolymer, and a charged H-P model \cite{10prlLiMS,13jcpLiMS} {\red{and multi-state Potts models \cite{schreck2010,zamparo2010,xiao2014}}} can be used to understand aggregation of proteins. The lattice dimer model \cite{Fowler} has been proposed to study the adsorption of diatomic molecules on a substrate. In this paper, we will use analytic equations to study finite-size  corrections and scaling  of the dimer model \cite{Fowler} on the checkerboard lattice.

Finite-size  corrections and scaling  for critical lattice systems \cite{Fowler,onsager,kasteleyn,fisher1961}, initiated more than four decades ago by Ferdinand, Fisher, and Barber \cite{ferdinand,FerdFisher,barber} have attracted much attention in recent decades (see Refs. \cite{privman,hu} for reviews). Finite-size effects become of practical interest due to the recent progress in fine processing technologies,  which has enabled the fabrication of nanoscale materials with novel shapes \cite{nano1,nano2,nano3}. In the quest to improve our understanding of realistic systems of finite extent, exactly solvable two-dimensional models play a key role in statistical mechanics  as they have long served as a testing ground to explore the general ideas of corrections and scaling under controlled conditions. Very few of them have been solved exactly and the dimer model \cite{Fowler,kasteleyn,fisher1961} is being one  of the most prominent examples.

The classical dimer model has been introduced in 1937 by Fowler and Rushbrook as a model for the adsorption of diatomic molecules on a substrate \cite{Fowler}. Later it became a general problem studied in various scientific communities with a large spectrum of applications. The dimer model has regained interest because of its quantum version, the so-called quantum dimer model, originally introduced by Rokhsar and Kivelson \cite{rokhsar}. Besides, a recent connection between dimer models and D-brane gauge theories has been discovered \cite{brane}, providing a very powerful computational tool.

\begin{figure}[tbp]
\includegraphics[width=0.32\textwidth]{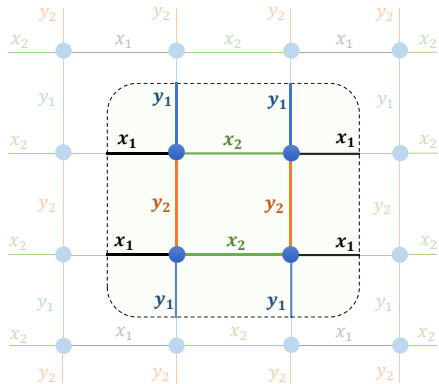}
\caption{(Color online) The unit cell for the dimer model on the checkerboard lattice.} \label{fig_1}
\end{figure}

>From the mathematical point of view the dimer model is extremely simple to define. We take a finite graph ${\mathcal L}$ and consider all arrangements of dimers (dominoes) so that all sites of ${\mathcal L}$ are covered by exactly one dimer. This is the so-called close-packed dimer model. Here we focus on the dimer model on a checkerboard lattice (see Fig. \ref{fig_1}). The checkerboard lattice is a unique two-dimensional (2D) system of great current interest, a set-up which provides a tool to study the evolution of physical properties as the system transits between different geometries. The checkerboard lattice is a simple rectangular lattice with anisotropic dimer weights $x_1, x_2, y_1$ and $y_2$. Each weight $a$ is simply the Boltzmann factor $e^{-E_a/k T}$ for a dimer on a bond of type $a$ with energy $E_a$. When one of the weights $x_1, x_2, y_1$, or $y_2$ on the checkerboard lattice is equal to zero, the partition function reduces to that for the dimer model on the one-dimensional strip. The dimer model on  the checkerboard lattice was first introduced by Kasteleyn \cite{kasteleyn1}, who showed that the model exhibits a phase transition.  The exact expression for the partition function for the dimer model on the checkerboard lattice on finite $2M \times 2N$ lattices with periodic boundary conditions has been obtained in Ref. \cite{ihk2015}.

In the present paper, we are going to study the finite-size effects of the dimer model on the finite checkerboard lattice of Fig. \ref{fig_1}.  The detailed study of the finite size effects for free energy and specific heat of the dimer model began with the work of Ferdinand \cite{ferdinand} few years after the exact solution, and has continued in a long series of articles using analytical \cite{blote,nagle,itzycson,izmailian2006,izmailian2007,wu2011,izmailian2014,izmkenna,ipph,ioh03,allegra,ruelle} and numerical  methods \cite{kong} for various geometries and boundary conditions. In particular, Ivashkevich, Izmailian, and Hu \cite{Ivasho}  proposed a systematic method to compute finite-size corrections to the partition functions and their derivatives of free models on torus,   including the Ising model, the dimer model, and the Gaussian model. Their approach is based on relations between the terms of the asymptotic expansion and the so-called Kronecker’s double series which are directly related to the elliptic $\theta$ functions.

We will apply the algorithm of Ivashkevich, Izmailian, and Hu \cite{Ivasho} to derive the exact asymptotic expansion of the logarithm  of the partition function. We will also derive the exact finite-size corrections for the free energy $F$, the internal energy  $U$, and the specific heat $C(t)$.  Using exact partition functions and finite-size corrections for the dimer model on the finite  checkerboard lattice we obtain finite-size scaling functions for  the free energy, the internal energy, and the specific heat. We are particularly interested in the finite-size scaling behavior of the specific-heat pseudocritical point. The pseudocritical point $t_{\mathrm{pseudo}}$ is the value of the temperature at which the specific heat has its maximum  for finite $2 M \times 2N$ lattice. One can determine this quantity as the point where the derivative of $C_{2M,2N}(t)$ vanishes. Finite-size properties of the specific heat $C_{2M,2N}(t)$ for the dimer model are characterized by (i) the location of its peak, $t_{\mathrm{pseudo}}$, (ii) its height $C(t_{\mathrm{pseudo}})$, and its value at the infinite-volume critical point $C(t_c)$. The peak position $t_{\rm{pseudo}}$, is a pseudocritical point which typically approaches $t_c$ as the characteristic size of the system $L$ tends to infinity as $L^{\lambda}$, where $\lambda$ is the shift exponent and $L$ is characteristic size of the system ($L=\sqrt{4 M N}$). Usually the shift exponent $\lambda$ coincides with $1/\nu$, where $\nu$ is the correlation length critical exponent, but this is not always the case and it is not a consequence of the finite-size scaling (FSS) \cite{barber1}. In a classic paper, Ferdinand and Fisher \cite{FerdFisher} determined the behavior of the specific heat pseudocritical point. They found that the shift exponent for the specific heat is $\lambda=1=1/\nu$, except for the special case of an infinitely long torus, in which case pseudocritical specific-heat scaling behavior was found to be of the form $L^2 \ln L$ \cite{onsager}. Thus the actual value of the shift exponent depends on the lattice topology (see Ref. \cite{kenna} and references therein). Quite recently Izmailian and Kenna \cite{izmkenna} have found that the shift exponent can be also depend on the parity of the number of lattice sites $N$ along a given lattice axis. They found for the dimer model on the triangular lattice that the shift exponent for the specific heat is equal to $1$ ($\lambda = 1$) for odd $N$, while for even $N$  the shift exponent is equal to infinite ($\lambda = \infty$). In the former case, therefore, the finite-size specific-heat pseudocritical point is size dependent, while in the latter case it coincides with the critical point of the thermodynamic limit. A question we wish to address here is the corresponding status of the shift exponent in the dimer model on the checkerboard lattice.

Our objective in this paper is to study the finite-size properties of a dimer model on the plane checkerboard lattice  using the same techniques developed in Refs. \cite{ioh03} and \cite{Ivasho}. The paper is organized as follows. In Sec. \ref{partition-function}  we introduce the dimer model on the checkerboard lattice with periodic boundary conditions. In Sec. \ref{asymptotic-expansion}  we derive the exact asymptotic expansions of the logarithm of the partition functions and their derivatives and write down the  expansion coefficients up to second order. In Sec. \ref{dimer-finitite-T} we numerically investigate the free energy, internal  energy and specific heat as function of temperature like parameter $t$, and analyze the scaling functions of the free energy, the internal energy,  and the specific heat. We also investigate the properties of the specific heat near the critical point and find  that the specific-heat shift exponent $\lambda$ is equal to infinity, which actually means that specific-heat pseudocritical point coincides with the critical point of the thermodynamic limit. In Sec. \ref{inf_long} we consider the limit $N \to \infty$ for which we obtain the expansion of the free energy for the dimer model on the infinitely long cylinder. From a finite-size analysis we find that the dimer model on a checkerboard lattice can be described by a conformal field theory having a central charge $c = - 2$. Our main results are summarized and discussed in Sec. \ref{summary-discussion}.

\section{Partition function}
\label{partition-function}
In the present paper, we consider the dimer model on an $2M \times 2N$ checkerboard lattice, as shown in Fig. \ref{fig_1}, under periodic boundary conditions. The partition function can be written as
\begin{equation}
Z=\sum x_1^{N_{x_1}}x_2^{N_{x_2}}y_1^{N_{y_1}}y_2^{N_{y_2}}, \label{partition}
\end{equation}
where $N_a$ is the number of dimers of type $a$ and the summation is over all possible dimer configurations on the lattice. An explicit expression for the partition function of the dimer model on the $2M \times 2N$ checkerboard lattice under periodic boundary condition is given by \cite{ihk2015}
\begin{eqnarray}
 Z_{2M,2N}(t)&=&\frac{(x_1x_2)^{M N}}{2}\left\{ -Z_{0,0}^2(t)+Z_{\frac{1}{2},\frac{1}{2}}^2(t)+
Z_{\frac{1}{2},0}^2(t)+Z_{0,\frac{1}{2}}^2(t)\right\},
\label{stat}\\
 Z_{\alpha,\beta}^2(t)&=&\prod_{m=0}^{M-1}\prod_{n=0}^{N-1} 4 \left\{t^2 +z^2
\sin^2 \left(\pi \frac{m+\beta}{M}\right)+ \sin^2 \left(
\pi\frac{n+\alpha}{N}\right) \right\} , \label{twist}
\end{eqnarray}
where
\begin{equation}
t^2 =\frac{(x_1-x_2)^2+(y_1-y_2)^2}{4x_1 x_2}, \qquad \qquad z^2 = \frac{y_1 y_2}{x_1 x_2}.
\end{equation}
Without loss the generality we can set $x_{1}x_{2}=1$ and $y_{1}y_{2}=1$, such that $z=1$.

The dimer model on checkerboard lattice has a singularity at $t=0$ ($x_1=x_2, y_1=y_2$). With the help of the identity
\begin{eqnarray}
 4\left|~\!{\sinh}\left(M\omega+i\pi\beta\right)\right|^2
=4\left[\,{\sinh}^2 M\omega +
\sin^2\pi\beta\,\right]=\prod_{m=0}^{M-1}4\textstyle{
\left\{~\!{\sinh}^2\omega +
\sin^2\left[\frac{\pi}{M}(m+\beta)\right]\right\}}, \label{ident}
\end{eqnarray}
the $Z_{\alpha,\beta}(t)$ can be transformed into a simpler form
\begin{eqnarray}
Z_{\alpha,\beta}(t)=\prod_{n=0}^{N-1} 2\left|
\textstyle{~\!{\sinh}\left\{M\omega_t\!\left(\pi\frac{n+\alpha}{N}\right)+i\pi \beta \right\} }\right|, \label{Zab}
\end{eqnarray}
where lattice dispersion relation has appeared
\begin{eqnarray}
\omega_t(k)={\rm arcsinh}\sqrt{\sin^2 k+t^2}.
\label{SpectralFunction}
\end{eqnarray}
The Taylor expansion of the lattice dispersion relation of Eq. (\ref{SpectralFunction}) at the critical point is given by
\begin{equation}
\omega_0(k)=k\left(\lambda_0+\sum_{p=1}^{\infty}
\frac{\lambda_{2p}}{(2p)!}\;k^{2p}\right),
\label{Spectral}
\end{equation}
where $\lambda_0=1$, $\lambda_2=-2/3$, $\lambda_4=4$, etc.

We are interested in computing the asymptotic expansions for large $M$, $N$ with fixed aspect ratio $\rho=M/N$ of the free energy $F_{2M,2N}(t)$, the internal energy $U_{2M,2N}(t)$, and the specific heat $C_{2M,2N}(t)$. These quantities are defined as follows
\begin{eqnarray}
F_{2M,2N}(t) &=& \frac{1}{4M N} \ln Z_{2M,2N}(t) \label{def_free_energy}, \\
U_{2M,2N}(t) &=& \frac{\partial}{\partial t} F_{2M,2N}(t) \label{def_energy},\\
C_{2M,2N}(t) &=&  \frac{\partial^2}{\partial t^2}F_{2M,2N}(t). \label{def_specific_heat}
\end{eqnarray}
In addition to $F_{2M,2N}(t)$, $U_{2M,2N}(t)$, and $C_{2M,2N}(t)$, we will also consider higher derivatives of the free energy at criticality
\begin{eqnarray}
F^{(k)}_c  =  \left.  \frac{\partial^k} {\partial t^k} F_{2M,2N}(t) \right|_{t=0},
\label{def_der_CH}
\end{eqnarray}
with $k=3,4$.

\section{Asymptotic expansion of the free energy and its derivatives}
\label{asymptotic-expansion}

\subsection{Asymptotic expansion of the free energy} \label{subsec:2a}
\label{subsec:2a0}
\begin{figure*}[tbp]
\includegraphics[width=0.36\textwidth]{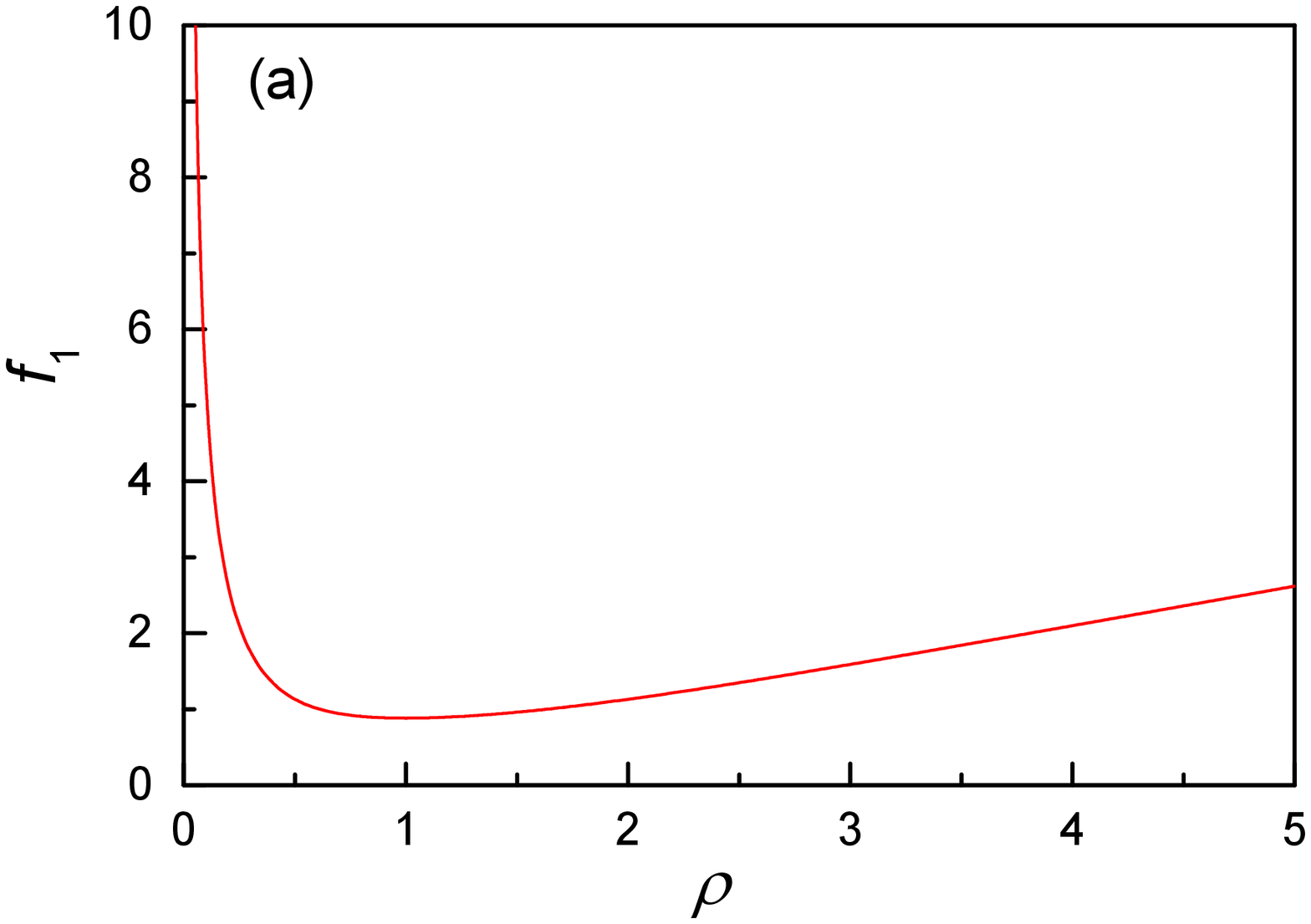}
\includegraphics[width=0.35\textwidth]{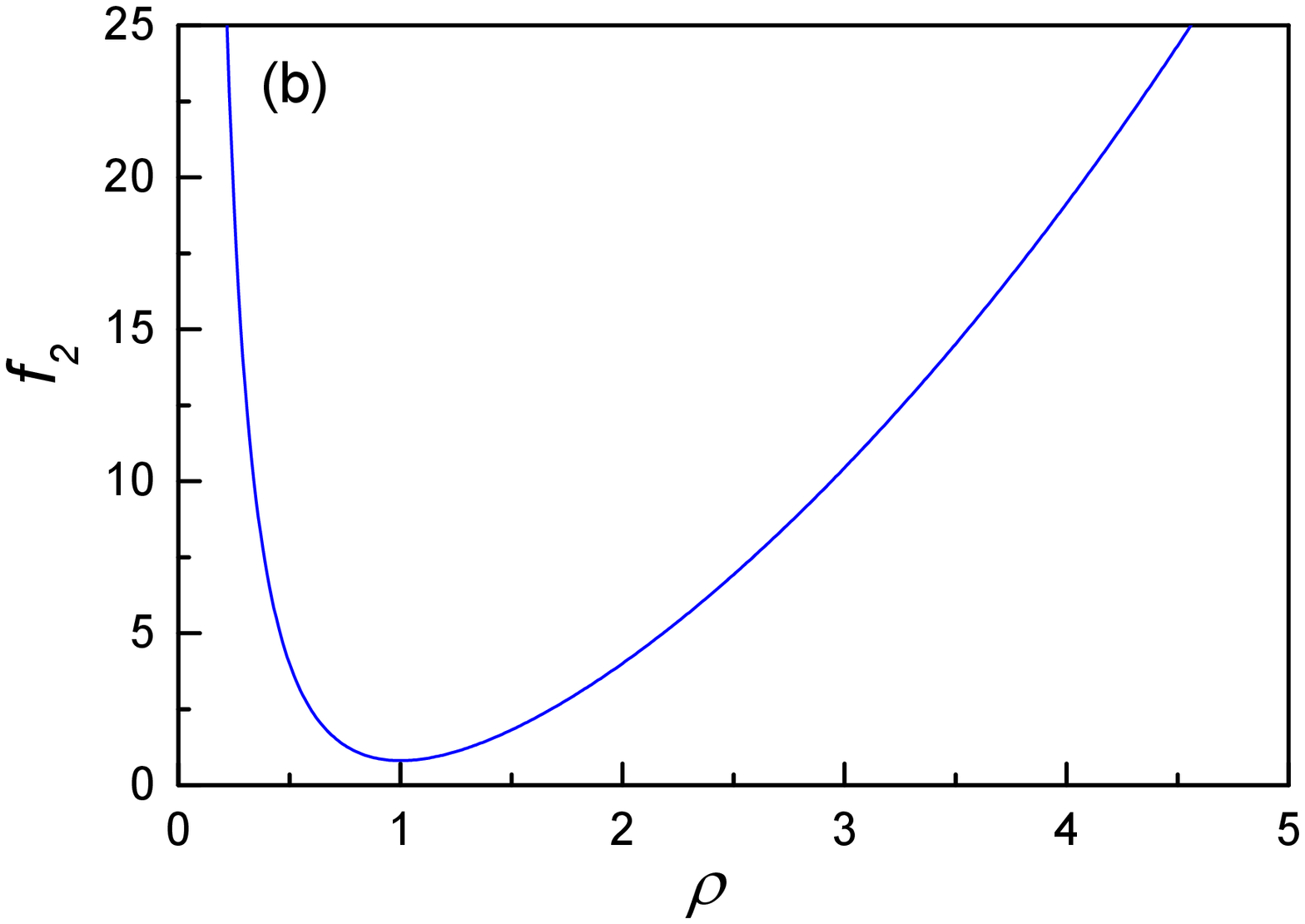}
\caption{(Color online) The values of the free energy asymptotic expansion coefficients (a) $f_1$ and (b) $f_2$ as functions of the aspect ratio $\rho$.}
\label{fig_2}
\end{figure*}

The exact asymptotic expansion of the logarithm of the partition function of the dimer model on checkerboard lattice at the critical point $t=t_c=0$ can be obtained along the same line as in Ref. \cite{Ivasho}. We didn't repeat here the calculations and give the final result:
\begin{equation}
\ln Z_{2M,2N}(0)=\ln\frac{1}{2}\left\{Z_{\frac{1}{2},\frac{1}{2}}^2(0)+
Z_{\frac{1}{2},0}^2(0)+Z_{0,\frac{1}{2}}^2(0)\right\}
\label{asymptotic}
\end{equation}
Here we use the fact that $Z_{0,0}^2(t)$ at the critical point is equal to zero. The exact asymptotic expansion of the $\ln Z_{\alpha,\beta}(0)$ for $(\alpha,\beta)= (0,\frac{1}{2}), (\frac{1}{2},0),(\frac{1}{2},\frac{1}{2})$ is given by \cite{Ivasho}
\begin{eqnarray}
\ln Z_{\alpha,\beta}(0)&=&\frac{S}{\pi}\int_{0}^{\pi}\!\!\omega_0(x)~\!{\rm
d}x + \ln\left|\frac{\theta_{\alpha,\beta}(i\lambda \rho)}{\eta(i\lambda \rho)}\right|
-2\pi\rho\sum_{p=1}^{\infty}
\left(\frac{\pi^2\rho}{S}\right)^{p}\frac{\Lambda_{2p}}{(2p)!}\,
\frac{{\tt Re}\;{\rm K}_{2p+2}^{\alpha,\beta}(i\lambda \rho)}{2p+2}
\label{ExpansionOflnZab}
\end{eqnarray}
Here $\theta_{\alpha,\beta}$ is the elliptic theta function ($\theta_{0,\frac{1}{2}}(i\lambda \rho)=\theta_2(i\lambda \rho)\equiv \theta_2, \theta_{\frac{1}{2},\frac{1}{2}}(i\lambda \rho)=\theta_3(i\lambda \rho)\equiv \theta_3, \theta_{\frac{1}{2},0}(i\lambda \rho)=\theta_4(i\lambda \rho)\equiv \theta_4$), $\eta(i\lambda \rho)\equiv \eta$ is the Dedekind $\eta$-function, ${\rm K}_{2p+2}^{\alpha,\beta}(i\lambda \rho)$ is the Kronecker’s double series (see Appendix D of Ref. \cite{Ivasho}), which can be expressed through the elliptic theta function (see Appendix F of Ref. \cite{Ivasho}) and $\Lambda_{2p}$ is the differential operators, which can be expressed via coefficients $\lambda_{2p}$ of the expansion of the lattice dispersion relation at the critical point as
\begin{eqnarray}
{\Lambda}_{2}&=&\lambda_2, \nonumber\\
{\Lambda}_{4}&=&\lambda_4+3\lambda_2^2\,\frac{\partial}{\partial\lambda},  \nonumber \\
{\Lambda}_{6}&=&\lambda_6+15\lambda_4\lambda_2\,\frac{\partial}{\partial\lambda}
+15\lambda_2^3\,\frac{\partial^2}{\partial\lambda^2}, \nonumber \\
&\vdots& \nonumber
\end{eqnarray}
Now using Eqs. (\ref{asymptotic}) and (\ref{ExpansionOflnZab}) it is easy to write down all terms in the exact asymptotic expansion of the logarithm of the partition function of the dimer model. Thus we find that the exact asymptotic expansion of the free energy at the critical point $F_c=F_{2M,2N}(0)$ can be written as
\begin{eqnarray}
F_c=F_{2M,2N}(0)=f_{\mathrm{bulk}} +\sum_{p=1}^\infty \frac{f_{p}(\rho)}{ S^{p}},
\label{expansion}
\end{eqnarray}
where $S = 4 M N$. The expansion coefficients are:
\begin{eqnarray}
f_{\mathrm{bulk}}&=&\frac{G}{\pi}=0.2915607\dots, \label{fbulk}\\
f_1(\rho) &=& \ln\frac{\theta_2^2+\theta_3^2+\theta_4^2}{2\eta^2}, \label{fex1}\\
f_2(\rho)&=&\frac{2\pi^3\rho^2}{45}\frac{\frac{7}{8}(\theta_2^{10}+\theta_3^{10}+\theta_4^{10})+\theta_2^2\theta_3^2
\theta_4^2(\theta_2^2 \theta_4^2- \theta_3^2\theta_2^2-\theta_3^2\theta_4^2)}{\theta_2^2+\theta_3^2+\theta_4^2}, \label{fex2}\\
& \vdots & \nonumber
\end{eqnarray}
where $G=0.915965\dots$ is the Catalan constant, and 
\begin{equation}
2\eta^3=\theta_2\theta_3\theta_4. \nonumber
\end{equation}
For the case when the aspect ratio $\rho$ is equal to $1$, the coefficients $f_1(\rho)$ and $f_2(\rho)$ are given by
\begin{equation}
f_1(\rho=1) = 0.881374\dots, \qquad f_2(\rho=1) = 0.805761\dots,
\label{aspectratio}
\end{equation}
which match very well with our numerical data (see Eq. (\ref{Fc})). The values of $f_1$ and $f_2$ as  functions of the aspect ratio $\rho$ are shown in Fig. \ref{fig_2}. {\red{Both of $f_1$ and $f_2$ have their minima at isotropic case, $\rho=1$.}}

\subsection{Asymptotic expansion of the internal energy}
\label{subsec:2b}
Now we will deal with the internal energy. The internal energy at the critical point can be computed directly from Eq. (\ref{def_energy}):
\begin{eqnarray}
U_c =  \frac{2}{S} \frac{-Z_{0,0}^\prime Z_{0,0}+Z_{0,1/2}^\prime Z_{0,1/2}+Z_{1/2,0}^\prime Z_{1/2,0}+Z_{1/2,1/2}^\prime Z_{1/2,1/2}}{-Z_{0,0}^2+Z_{0,1/2}^2+Z_{1/2,0}^2+Z_{1/2,1/2}^2}, \label{critical_energy}
\end{eqnarray}
Here $Z_{\alpha,\beta}^{\prime}=\left. \frac{d Z_{\alpha,\beta}(t)}{d t} \right|_{t=0}$ is the first derivative of $Z_{\alpha,\beta}(t)$ with respect to $t$ at criticality. In what follow we will use the following notation
\begin{equation}
\left. Z_{\alpha,\beta}(t)\right|_{t=0}=Z_{\alpha,\beta}, \left. Z_{\alpha,\beta}(t)^{\prime}\right|_{t=0}=Z_{\alpha,\beta}^{\prime}, \left. Z_{\alpha,\beta}(t)^{\prime \prime}\right|_{t=0}=Z_{\alpha,\beta}^{\prime \prime}, \left. Z_{\alpha,\beta}(t)^{\prime \prime \prime}\right|_{t=0}=Z_{\alpha,\beta}^{\prime \prime \prime},
\left. Z_{\alpha,\beta}(t)^{(4)}\right|_{t=0}=Z_{\alpha,\beta}^{(4)}.
\end{equation}
Since
\begin{equation}
Z_{0,0}=Z_{0,1/2}^\prime=Z_{1/2,0}^\prime=Z_{1/2,1/2}^\prime=0,
\end{equation}
the internal energy at the critical point is equal to zero
\begin{equation}
U_c=0.
\end{equation}

\subsection{Asymptotic expansion of the specific heat}
\label{subsec:2c}
\begin{figure*}[tbp]
\includegraphics[width=0.36\textwidth]{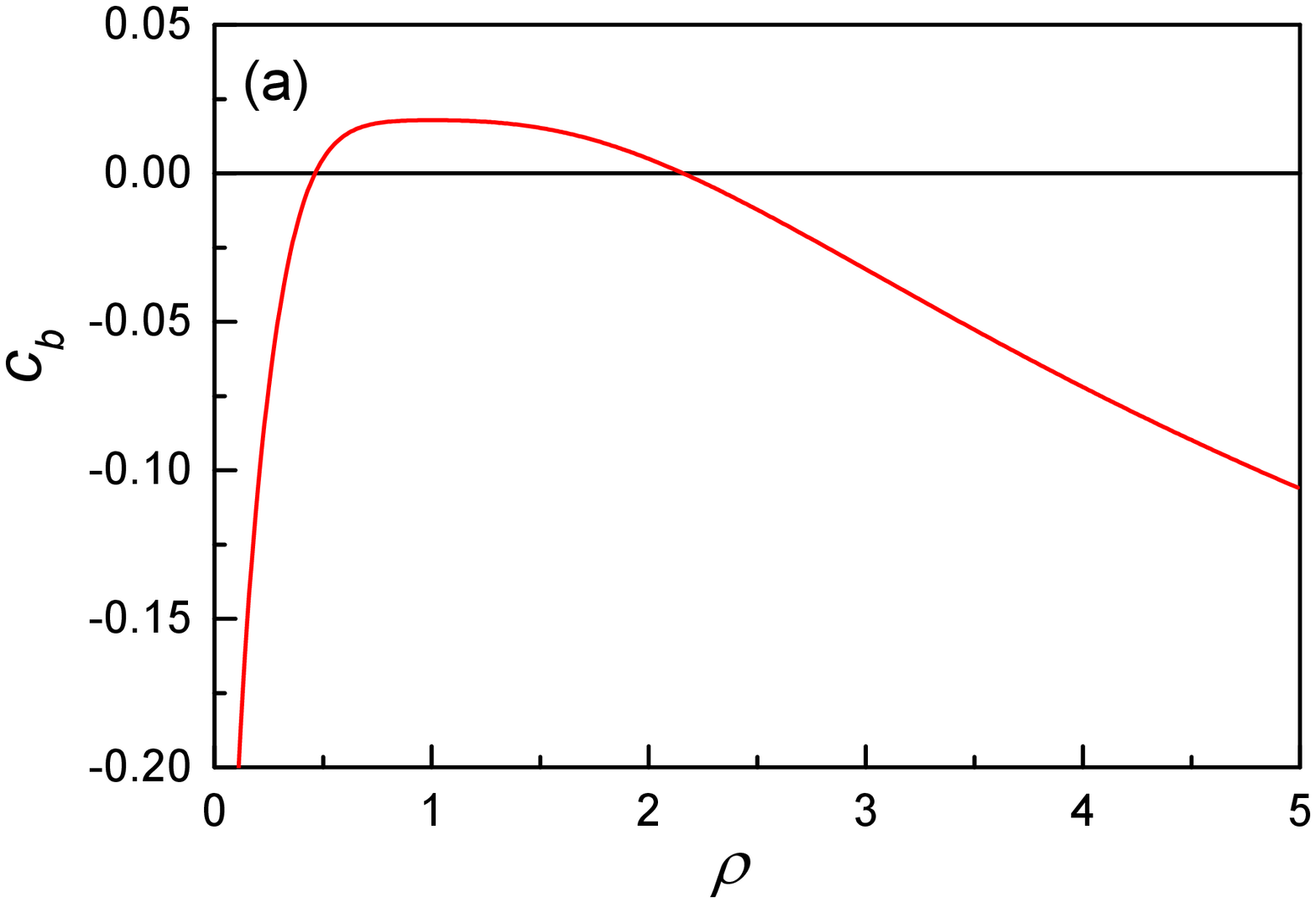}
\includegraphics[width=0.35\textwidth]{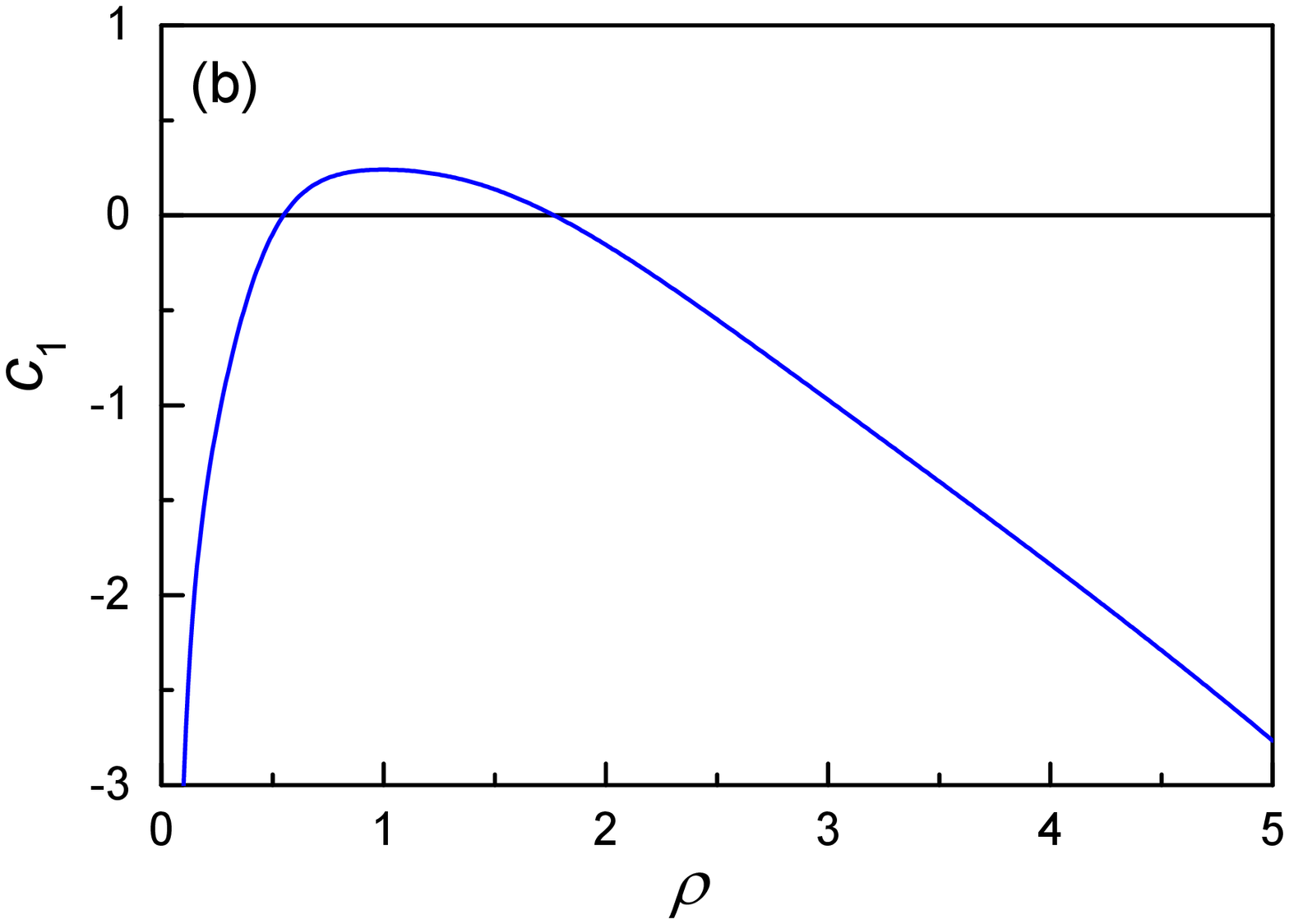}
\caption{(Color online) The values of the specific heat asymptotic expansion coefficients (a) $c_b$ and (b) $c_1$ as functions of the aspect ratio $\rho$.}
\label{fig_3}
\end{figure*}

The specific heat at criticality is given by the following formula
\begin{eqnarray}
 C_c  &=& \frac{2}{S}
            \frac{-Z_{0,0}^{\prime^2}+Z_{0,1/2}Z_{0,1/2}^{\prime \prime}+Z_{1/2,0}Z_{1/2,0}^{\prime \prime}+Z_{1/2,1/2}Z_{1/2,1/2}^{\prime \prime}}
            {Z_{0,1/2}^2+Z_{1/2,0}^2+Z_{1/2,1/2}^2}.
\label{def_CH}
\end{eqnarray}
Following along the same line as in Ref. \cite{Ivasho} we have found that the exact asymptotic expansion of the specific heat can be written in the following form
\begin{eqnarray}
C_c &=& \frac{1}{2\pi}\ln{S}+ c_{b}+\sum_{p=1}^{\infty} \frac{c_{p}}{S^{p}}+ \cdots \nonumber \\
&=& \frac{1}{2\pi}\ln{S}+c_{b}+ \frac{c_{1}}{S}+\cdots . \label{heatexpansion}
\end{eqnarray}
where
\begin{eqnarray}
 c_{b} &=& \frac{1}{\pi}
\left(C_E-\frac{1}{2}\ln{\rho}-\ln{\pi}+\frac{3}{2}\ln{2}\right)
- \frac{\rho}{2} \frac{
{\theta}_2^2{\theta}_3^2{\theta}_4^2}{{\theta}_2^2+{\theta}_3^2+{\theta}_4^2}
- \frac{2}{\pi}\frac{\sum_{i=2}^4{\theta}_i^2 \ln{{\theta}_i}}
{{\theta}_2^2+{\theta}_3^2+{\theta}_4^2},
\label{cbulk} \\
 c_1 &=& \frac{\pi^2 \rho^2}{6}~\frac{{\theta}_2^2{\theta}_3^2{\theta}_4^2}{({\theta}_2^2+{\theta}_3^2+{\theta}_4^2)^2}
 \left[{\theta}_4^2({\theta}_2^4+{\theta}_3^4)\ln\frac{{\theta}_2}{{\theta}_3}+{\theta}_3^2({\theta}_2^4
 -{\theta}_4^4)\ln\frac{{\theta}_4}{{\theta}_2}+{\theta}_2^2({\theta}_3^4+{\theta}_4^4)\ln\frac{{\theta}_4}{{\theta}_3}\right] \nonumber \\
&+& \frac{\pi^2 \rho^2}{18}~ \frac{{\theta}_3^4 {\theta}_4^4
(2{\theta}_2^2
-{\theta}_3^2-{\theta}_4^2)}{{\theta}_2^2+{\theta}_3^2+{\theta}_4^2}
+\frac{\pi^3 \rho^3}{24} \frac{{\theta}_2^2 {\theta}_3^2
{\theta}_4^2 ({\theta}_2^{10}+{\theta}_3^{10}+{\theta}_4^{10})}
{({\theta}_2^2+{\theta}_3^2+{\theta}_4^2)^2} \nonumber \\
&+& \frac{\pi \rho}{18}~ \frac{(\theta_2^2-\theta_4^2)({\theta}_3^4-\theta_2^2{\theta}_4^2+
{\theta}_2^2{\theta}_3^2+{\theta}_3^2{\theta}_4^2)}{{\theta}_2^2+{\theta}_3^2+{\theta}_4^2}
\left(1+4 ~\rho\;\frac{\partial}{\partial \rho}
\ln{\theta_2}\right). 
\label{c1}
\end{eqnarray}
Here $C_{E}=0.5772156649\dots$ is the Euler constant and
\begin{equation}
\frac{\partial}{\partial \rho}
\ln{\theta_2}=-\frac{1}{2}\theta_3^2 E,
\end{equation}
where $E$ is the elliptic integral of the second kind. Note that the $c_b$ and $c_1$ are functions of the aspect ratio $\rho$. For the case when the aspect ratio $\rho$ is equal to $1$ the coefficients $c_b$ and $c_1$ are given by
\begin{equation}
c_{b} (\rho=1) = 0.0178829\dots, \qquad c_{1} (\rho=1)=0.240428\dots,
\label{aspectratio1}
\end{equation}
which match very well with our numerical data (see Eq. (\ref{cmax_scaling})). The values of $c_b$ and $c_1$ as functions of $\rho$ are shown in Fig. \ref{fig_3}. {\red{Interestingly, the non-monotonicity $c_b$ and $c_1$ as functions of anisotropy have maximum values at $\rho=1$.}}

\subsection{Asymptotic expansion of the higher derivatives of the free energy}
\label{higher}
Using the fact that
\begin{equation}
Z_{0,0}=Z_{0,1/2}^\prime=Z_{1/2,0}^\prime=Z_{1/2,1/2}^\prime=Z_{0,0}^{\prime \prime}=Z_{0,1/2}^{\prime \prime\prime}=Z_{1/2,0}^{\prime\prime \prime}=Z_{1/2,1/2}^{\prime\prime \prime}=0,
\end{equation}
it is easy to show that the third derivative of the logarithm of the partition function at the criticality $F^{(3)}_c$ is equal to zero
\begin{eqnarray}
 F^{(3)}_c &=&0. 
\end{eqnarray}
Let us now  consider the fourth derivative of the logarithm of the partition function at the criticality $F^{(4)}_c$ which can be written as follows:
\begin{eqnarray}
 F^{(4)}_c &=& -3 S\; C_c^2
 +\frac{6}{S}\frac{Z_{0,1/2}^{\prime \prime^2}+Z_{1/2,0}^{\prime \prime^2}+Z_{1/2,1/2}^{\prime \prime^2}}
 {Z_{0,1/2}^2+Z_{1/2,0}^2+Z_{1/2,1/2}^2}-\frac{8}{S}\frac{Z_{0,0}^{\prime} Z_{0,0}^{\prime \prime \prime} } {Z_{0,1/2}^2+Z_{1/2,0}^2+Z_{1/2,1/2}^2} \nonumber \\
 &+& \frac{2}{S}\frac{Z_{0,1/2} Z_{0,1/2}^{(4)}+Z_{1/2,0}Z_{1/2,0}^{(4)}+Z_{1/2,1/2} Z_{1/2,1/2}^{(4)}}
 {Z_{0,1/2}^2+Z_{1/2,0}^2+Z_{1/2,1/2}^2}.
 \label{def_F4critchislo}
\end{eqnarray}
We have found that the exact asymptotic expansion  can be written in the following form
\begin{eqnarray}
F_c^{(4)} &=& g S -\frac{3}{2\pi} \ln{S}+g_0 +\sum_{p=1}^{\infty}
\frac{g_{p}}{S^{p}} \nonumber \\
&=& g S -\frac{3}{2\pi}\ln{S} +g_{0}+\frac{g_{1}}{S}+\cdots , \label{freeenergy4expansion}
\end{eqnarray}
where
\begin{eqnarray}
g(\rho) &=& \frac{12}{\pi^2}\;\frac{
 {\theta}_3^2{\theta}_4^2\left(\ln{\frac{{\theta}_3}{{\theta}_4}}\right)^2
 +{\theta}_2^2{\theta}_4^2\left(\ln{\frac{{\theta}_2}{{\theta}_4}}\right)^2
 +{\theta}_2^2{\theta}_3^2\left(\ln{\frac{{\theta}_2}{{\theta}_3}}\right)^2}
 {({\theta}_2^2+{\theta}_3^2+{\theta}_4^2)^2}
\nonumber \\
&-& \frac{3\rho}{4} \frac{
{\theta}_2^2{\theta}_3^2{\theta}_4^2}{{\theta}_2^2+{\theta}_3^2+{\theta}_4^2}
 \left[ \rho \frac{
{\theta}_2^2{\theta}_3^2{\theta}_4^2}{{\theta}_2^2+{\theta}_3^2+{\theta}_4^2}+\frac{8}{\pi}
\left(\frac{\sum_{i=2}^4{\theta}_i^2 \ln{{\theta}_i}}
{{\theta}_2^2+{\theta}_3^2+{\theta}_4^2}-\ln{2 \eta}\right)\right]
\nonumber \\
&+& \frac{3}{16\pi^3 \rho}\;\frac{{\theta}_2^2
\left(\rho\frac{\partial}{\partial \rho }-1\right)R_4^{0,1/2}(\rho)+{\theta}_3^2 \left(\rho\frac{\partial}{\partial \rho
}-1\right)R_4^{1/2,1/2}(\rho)+{\theta}_4^2
\left(\rho\frac{\partial}{\partial \rho }-1\right)R_4^{1/2,0}(\rho)}{{\theta}_2^2+{\theta}_3^2+{\theta}_4^2}.
\label{sm3}
\end{eqnarray}
$R_4^{\alpha,\beta}$ is given by
\begin{eqnarray}
R_4^{\alpha,\beta}(\rho) = -\psi^{\prime \prime}(\alpha)-\psi^{\prime \prime}(1-\alpha)+4\sum_{n=0}^{\infty}\sum_{m=1}^{\infty}
\left\{\frac{e^{-2\pi m(\rho(n+\alpha)+i \beta)}}{(n+\alpha)^3}+(\alpha \to 1-\alpha)\right\},
\end{eqnarray}
where $\psi^{\prime \prime}(x)$ is the second derivative of the digamma function $\psi(x)$ with respect to $x$,
\begin{equation}
\psi^{\prime\prime}(1) = -2\zeta(3), \qquad \psi^{\prime\prime}(1/2)=-14\zeta(3).
\end{equation}
Here $\zeta(n)$ is the zeta function
\begin{equation}
\zeta(n) = \sum_{k=1}^{\infty}\frac{1}{k^n},
\end{equation}
and for small $x$
\begin{eqnarray}
\psi(x) &=& -C_E-\frac{1}{x} +x\sum_{k=1}^{\infty}\frac{1}{k(k+x)}, \nonumber \\
\psi^{\prime \prime}(x)&=&-\frac{2}{x^3}+2x\sum_{k=1}^{\infty}\frac{1}{k(k+x)^3}-2\sum_{k=1}^{\infty}\frac{1}{k(k+x)^2}. \nonumber
\end{eqnarray}

\begin{figure}[tbp]
\includegraphics[width=0.40\textwidth]{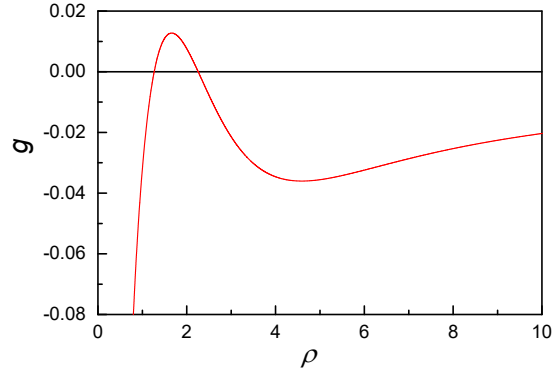}
\caption{(Color online) The value of the asymptotic expansion coefficient $g$ in the fourth derivative of free energy $F_c^{(4)}$ as a function of the aspect ratio $\rho$.}
\label{fig_4}
\end{figure}

One can show that
\begin{eqnarray}
\left(\rho\frac{\partial}{\partial \rho }-1\right)R_4^{0,\frac{1}{2}}(\rho)&=&-\frac{4\pi^3\rho}{3}-4\zeta(3)+\sum_{n=1}^{\infty}\frac{8}{n^3\left(1+e^{2\pi n\rho}\right)}+\sum_{n=1}^{\infty}\frac{4\pi\rho}{n^2\cosh^2(\pi n\rho)}, \nonumber \\
\left(\rho\frac{\partial}{\partial \rho }-1\right)R_4^{\frac{1}{2},\frac{1}{2}}(\rho)&=&-28\zeta(3)+\sum_{n=1}^{\infty}\frac{8}{\left(n+\frac{1}{2}\right)^3
\left(e^{2\pi\rho\left(n+\frac{1}{2}\right)}+1\right)}+
\sum_{n=1}^{\infty}\frac{4\pi\rho}{\left(n+\frac{1}{2}\right)^2\cosh^2{(\pi\rho\left(n+\frac{1}{2})\right)}}, \nonumber \\
\left(\rho\frac{\partial}{\partial \rho }-1\right)R_4^{\frac{1}{2},0}(\rho)&=&-28\zeta(3)-\sum_{n=1}^{\infty}\frac{8}{\left(n+\frac{1}{2}\right)^3
\left(e^{2\pi\rho\left(n+\frac{1}{2}\right)}-1\right)}-
\sum_{n=1}^{\infty}\frac{4\pi\rho}{\left(n+\frac{1}{2}\right)^2\sinh^2{(\pi\rho\left(n+\frac{1}{2})\right)}}. \nonumber
\end{eqnarray}
Accordingly, the value of the asymptotic expansion coefficient $g$ of Eq.(\ref{sm3}) as a function of the aspect ratio $\rho$ can be determined and is shown in Fig. \ref{fig_4}. More explicitly, $g(\rho=1)=-0.032122\dots$, $g(\rho=2)=0.00762119\dots$, and  $g(\rho=4)=-0.0346017\dots$. {\red{The maximum of $g(\rho)$ takes place at $\rho=5/3$.}}

\section{Dimer model on the checkerboard lattice at finite temperature}
\label{dimer-finitite-T}

\subsection{Numerical calculations of thermodynamic variables}
\begin{figure}[tbp]
\includegraphics[width=0.32\textwidth]{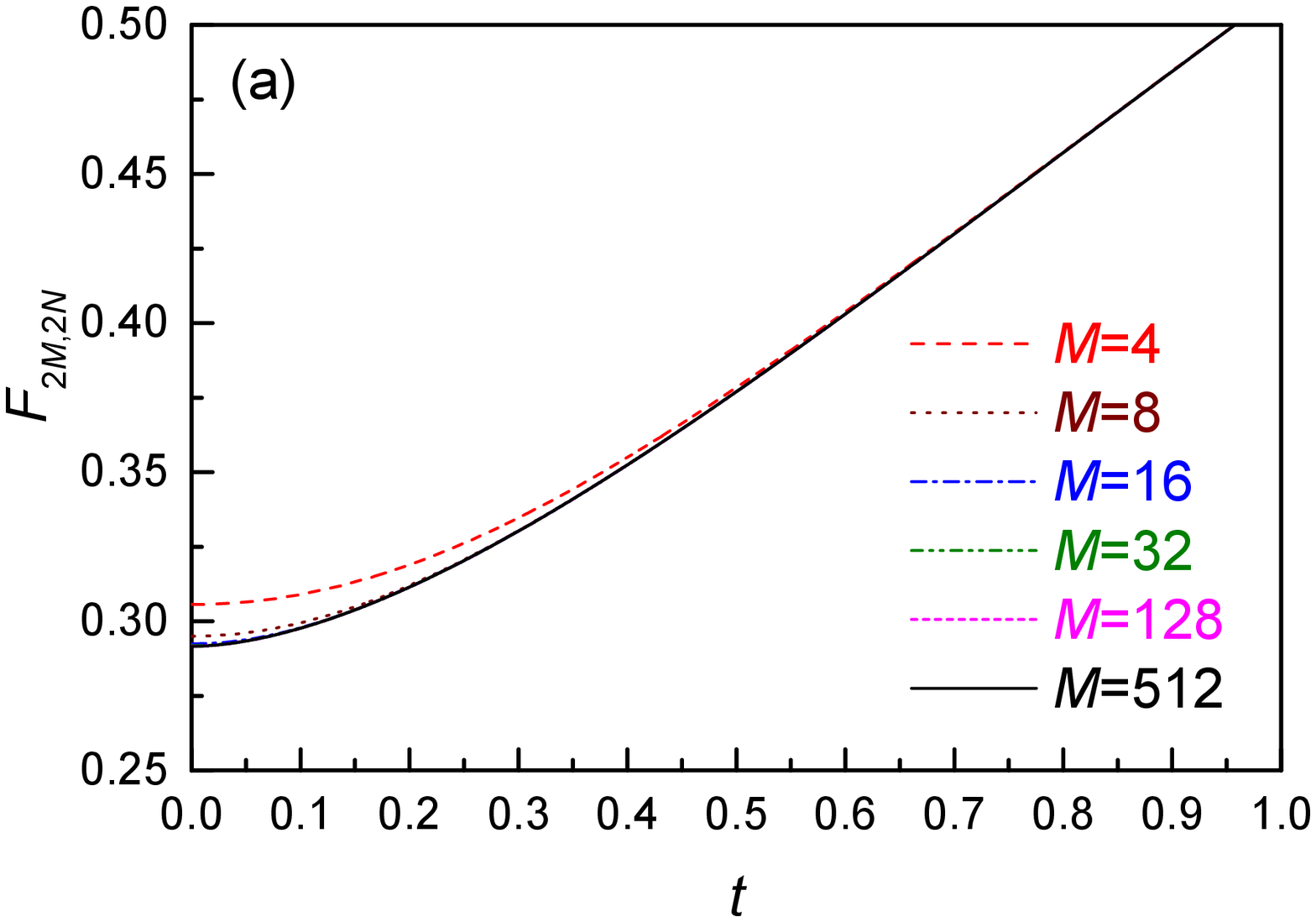}
\includegraphics[width=0.32\textwidth]{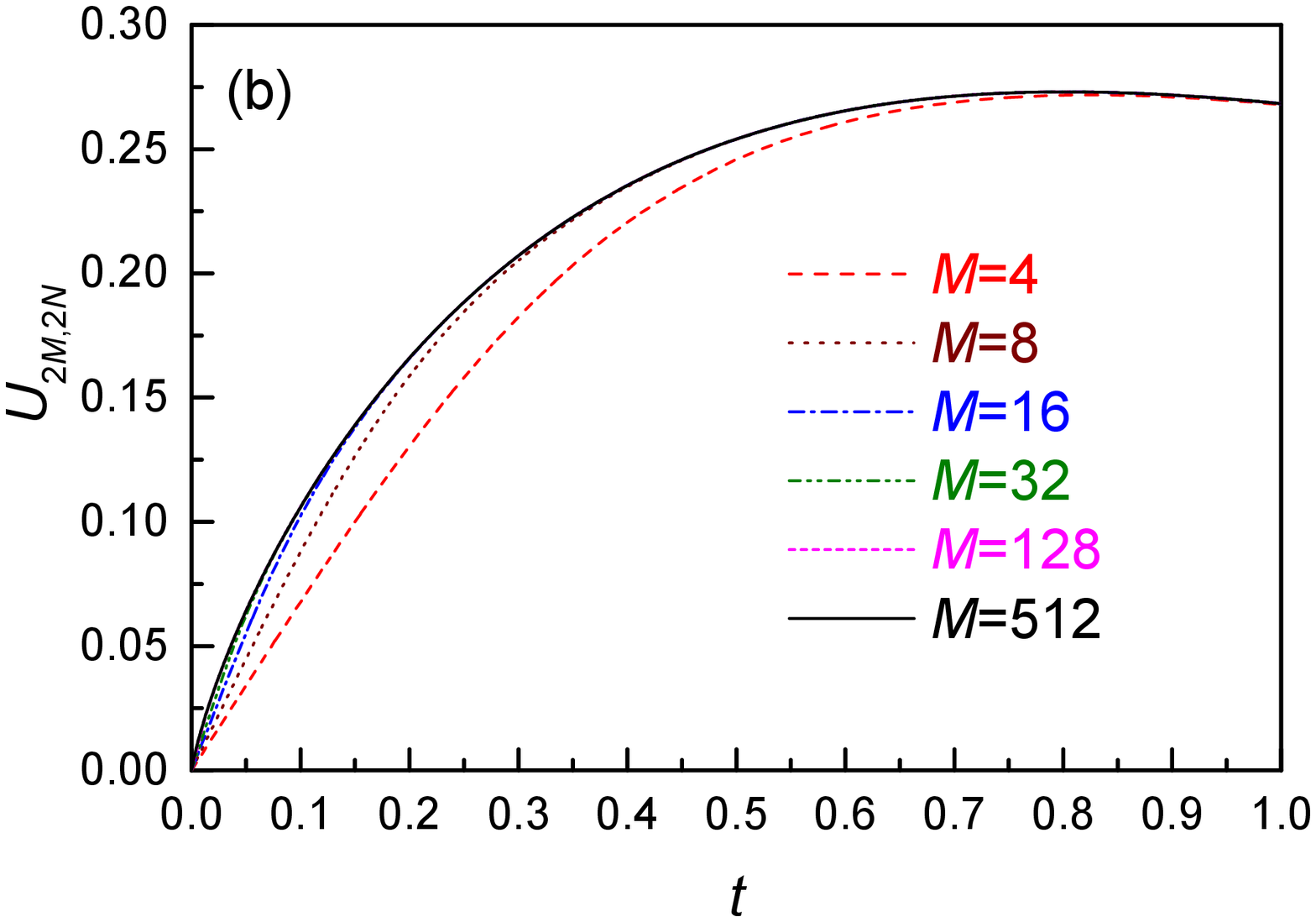}
\includegraphics[width=0.32\textwidth]{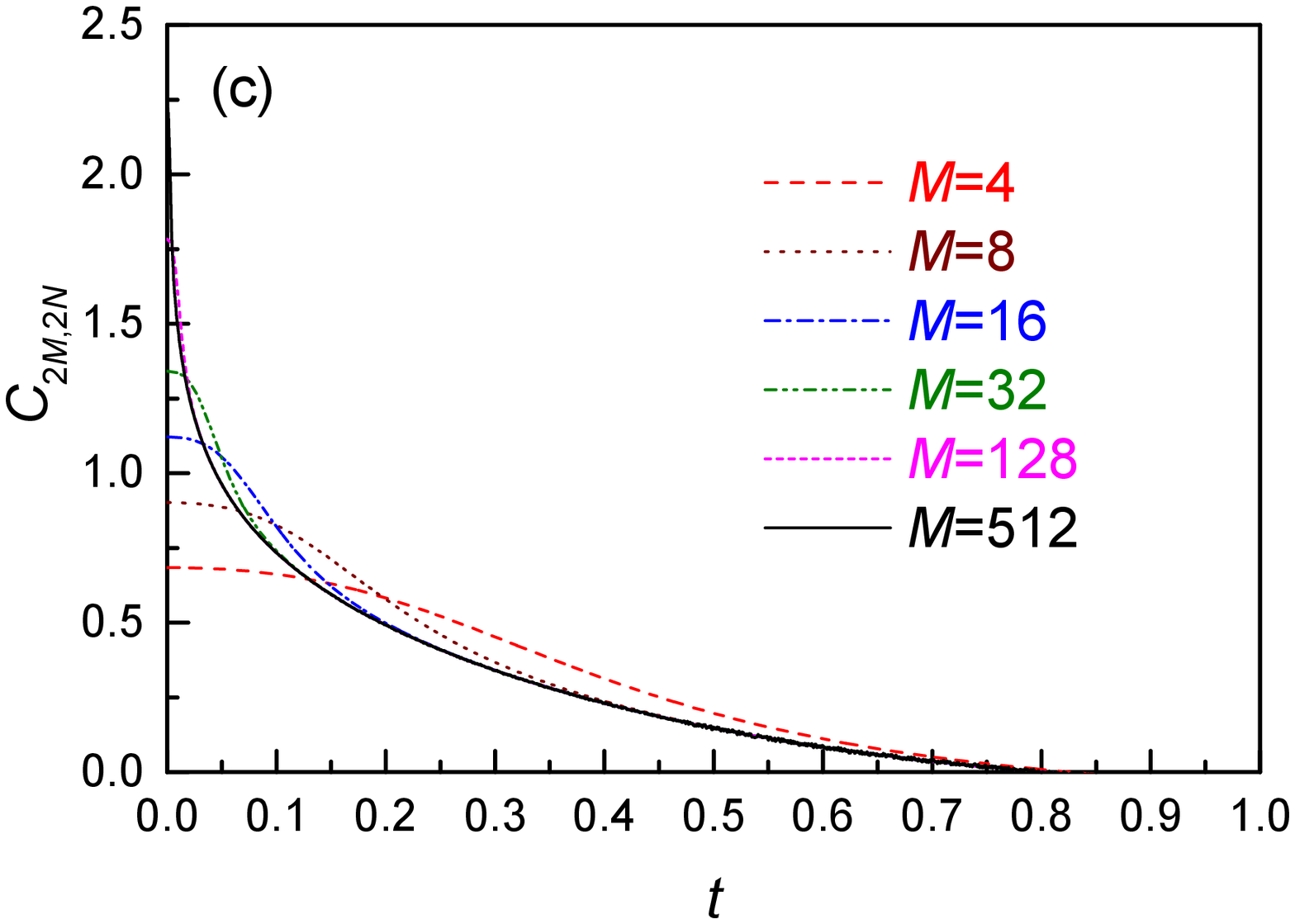}
\caption{(Color online) (a) Free energy $F_{2M,2N}(t)$, (b) internal energy $U_{2M,2N}(t)$ and (c) specific heat $C_{2M,2N}(t)$ as functions of $t$. The aspect ratio $\rho=M/N$ has been set to unity.} \label{fig_5}
\end{figure}

\begin{figure}[tbp]
\includegraphics[width=0.42\textwidth]{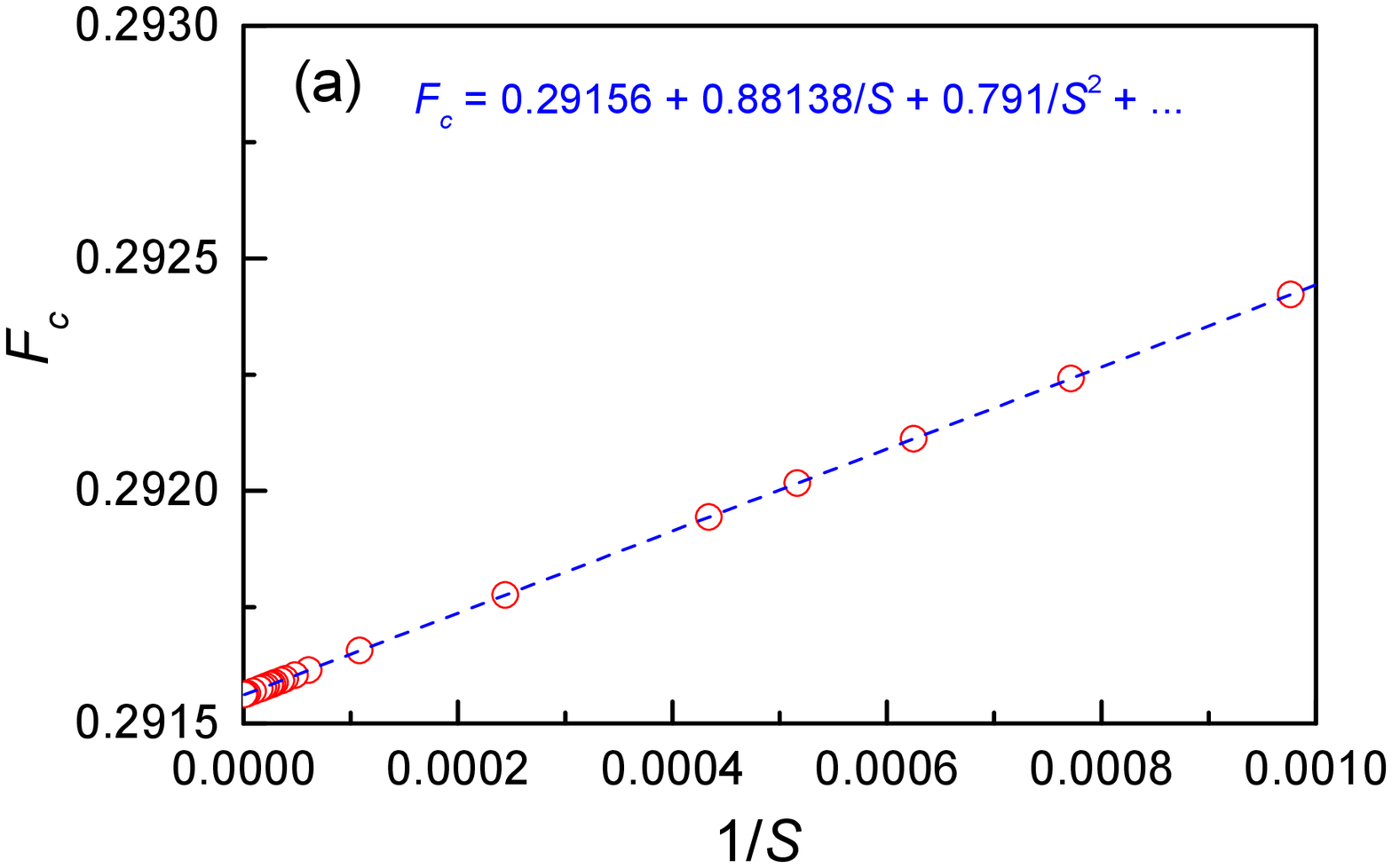}
\includegraphics[width=0.42\textwidth]{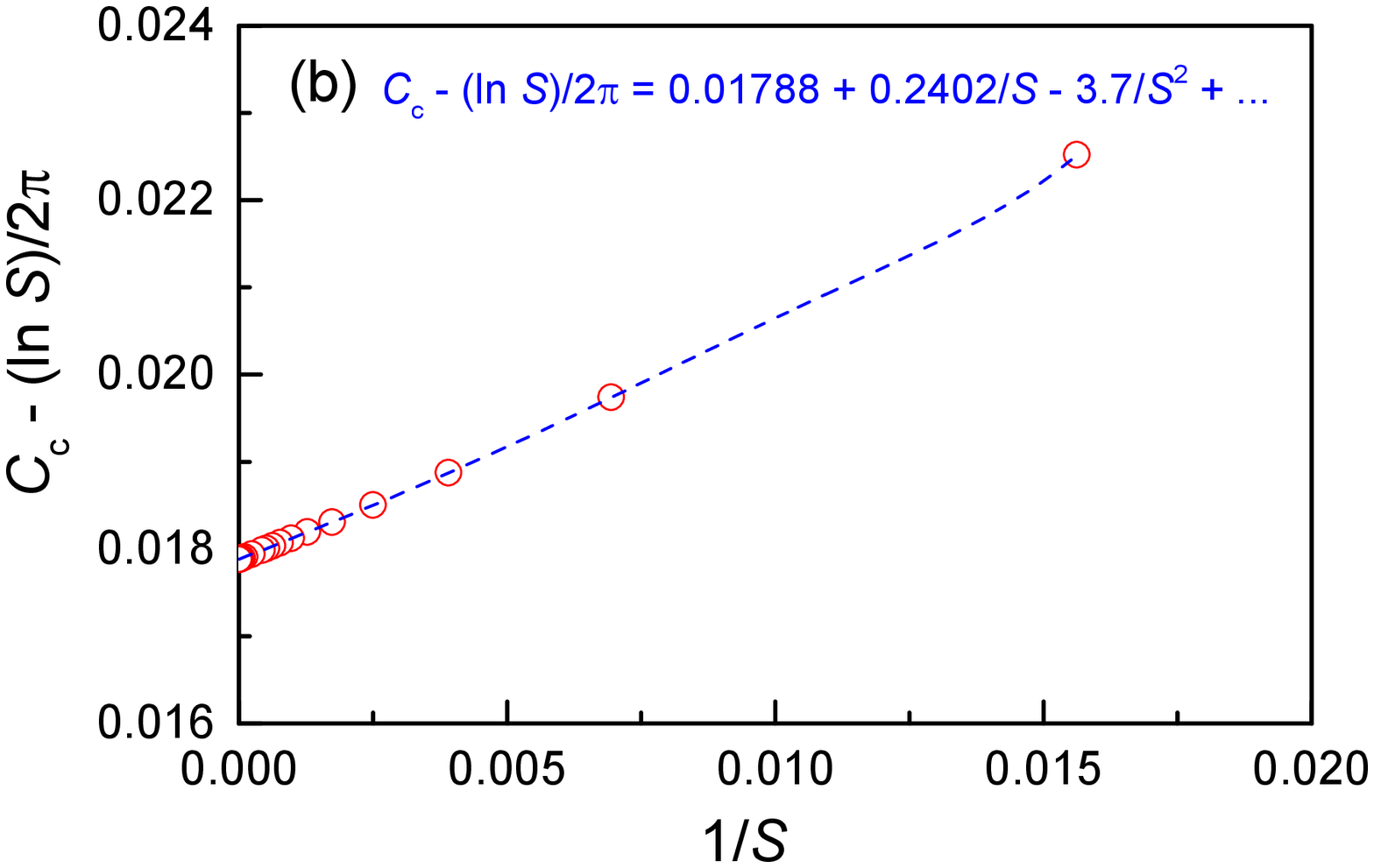}
\caption{(Color online) (a) $F_{\mathrm{c}}$ as a function of $1/S$, and (b) $C_{\mathrm{c}}-(\ln S)/2\pi $ as functions of $1/S$. The aspect ratio $\rho=M/N$ has been set to unity.}
\label{fig_6}
\end{figure}

Using the partition function of Eq. (\ref{stat}) we plot the free energy $F_{2M,2N}(t)$, the internal energy $U_{2M,2N}(t)$ and the specific heat $C_{2M,2N}(t)$ as functions of $t$  for different lattice sizes in Figs. \ref{fig_5}(a), \ref{fig_5}(b), and \ref{fig_5}(c), respectively.
The specific heat curve becomes higher with the increase of the system size, while the peaks always locate at $t=0$ exactly. To study scaling behaviors of thermodynamic variables, we analyzed the variation of $F_{\mathrm{c}}=F(t_{\mathrm{c}})$ with respect to different system sizes $S=2M \times 2N$. Figure \ref{fig_6}(a) shows $F_{\mathrm{c}}$ as a function of $1/S$. Using a polynomial function of $1/S$ to fit the data, the best polynomial fitting to the data is found to be
\begin{equation}
F_{\mathrm{c}} = 0.29156 (\pm 0.00000001) + \frac {0.88138 (\pm 0.00001)}{S} + \frac{0.791 (\pm 0.004)}{S^2} + \cdots,
\end{equation}
which can be approximately expressed as
\begin{equation}
F_{\mathrm{c}} \approx  \frac{G}{\pi} + \frac {0.88138 (\pm 0.00001)}{S} + \frac{0.791 (\pm 0.004)}{S^2}. \label{Fc}
\end{equation}
This expression is consistent with Eq. (\ref{expansion}) for the case $\rho = 1$, see Eq. (\ref{aspectratio}). For the specific heat, we plotted $C_{\mathrm{c}} - \ln S/(2\pi)$ as a function of $1/S$ in Fig. \ref{fig_6}(b). The data points can be well described by the polynomial fit
\begin{equation}
C_{\mathrm{c}} - \frac{1}{2\pi} \ln S =  0.01788 (\pm 0.000000009) + \frac{0.2402 (\pm 0.001)}{S} - \frac{3.7 (\pm 1.9)}{S^2} \cdots.
\label{cmax_scaling}
\end{equation}
This expression is also consistent with Eq.(\ref{heatexpansion}) for the case $\rho = 1$, see Eq. (\ref{aspectratio1}).

\subsection{Scaling functions of the free energy, the internal energy and the specific heat}
\label{scaling-functions}
\begin{figure}[tbp]
\includegraphics[width=0.32\textwidth]{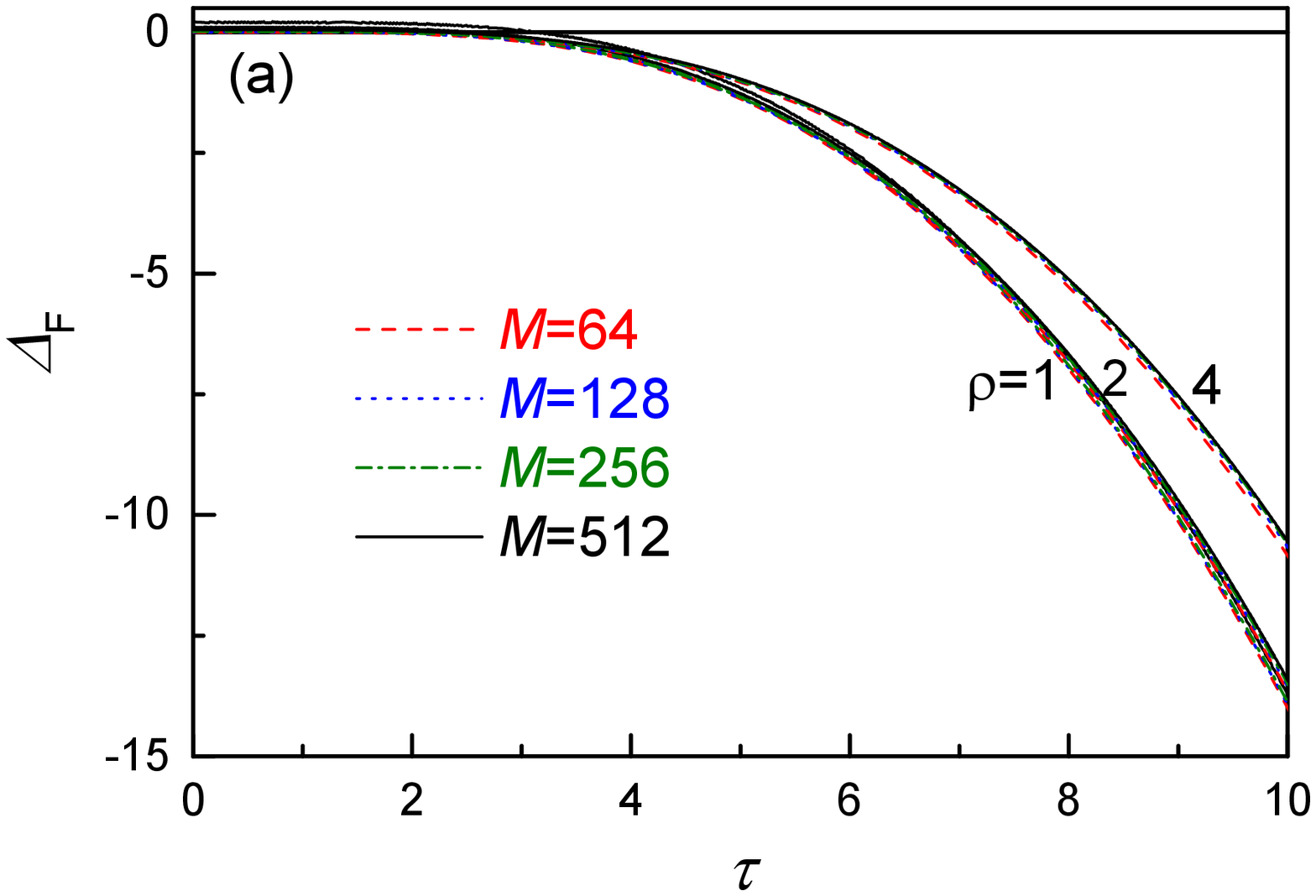}
\includegraphics[width=0.32\textwidth]{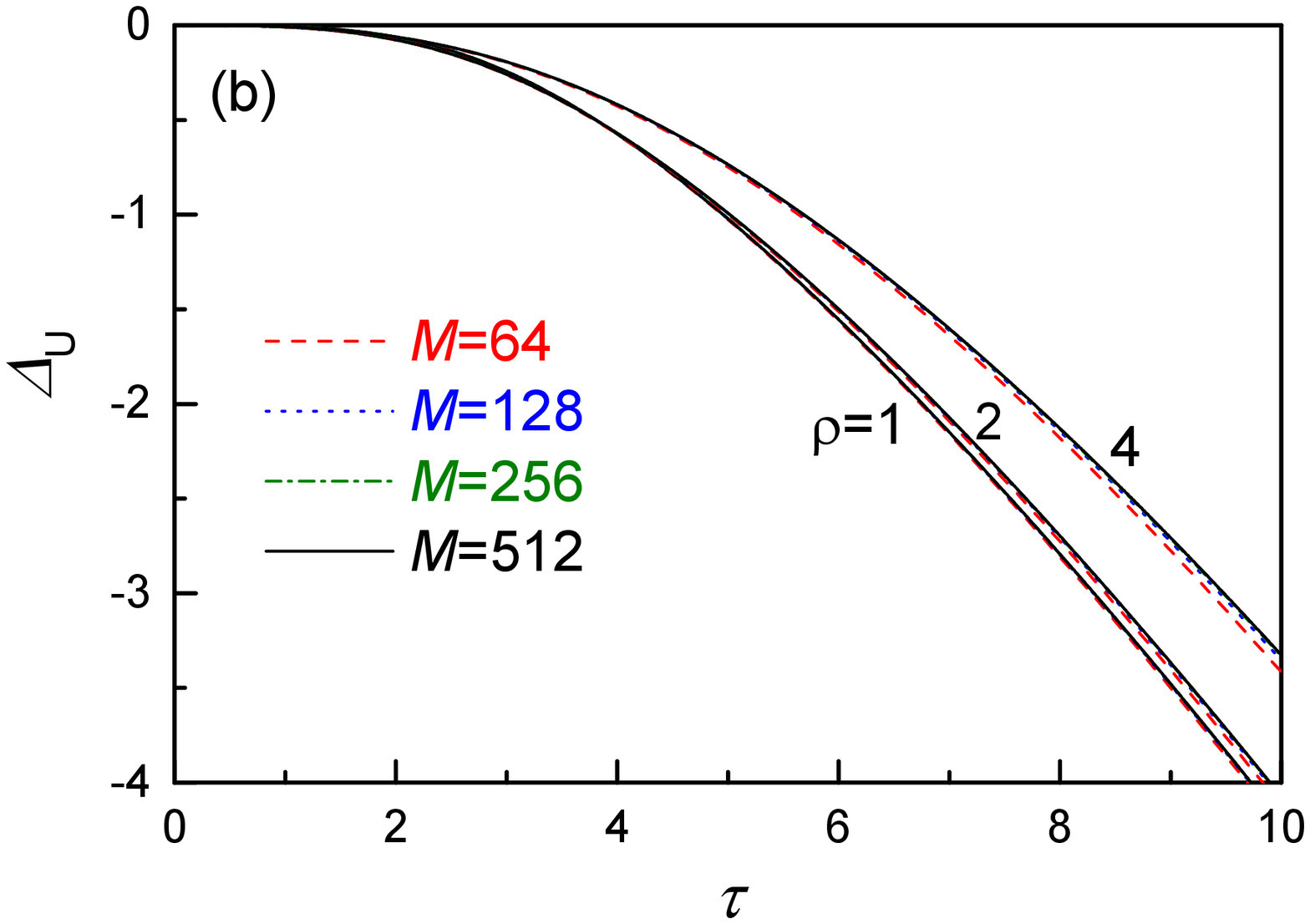}
\includegraphics[width=0.33\textwidth]{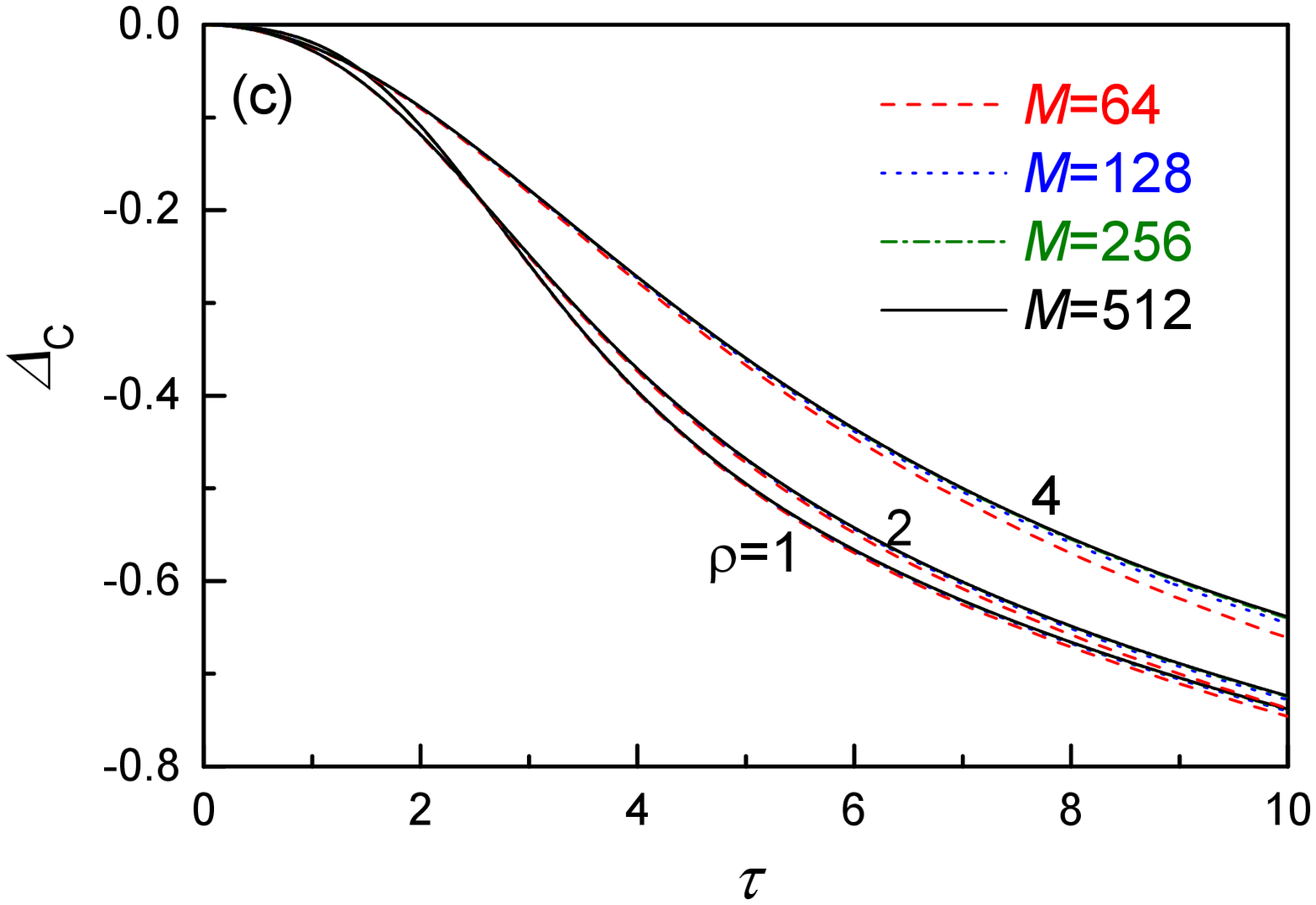}
\caption{(Color online) The scaling functions (a) $\Delta _F (S, \rho, \tau)$, (b) $\Delta _U (S, \rho, \tau)$
and (c) $\Delta _C (S, \rho, \tau)$, as functions of $\tau$ for different system sizes $S$ with the aspect ratio $\rho=1$, $2$, and $4$.}
\label{fig_7}
\end{figure}

Following the proposal for the scaling functions in Ref. \cite{wu2003} and using the exact expansions of the free energy in Eq. (\ref{expansion}) and the specific heat in Eq. (\ref{heatexpansion}), we define the scaling function of the free energy $\Delta _F (S, \rho, t)$ as
\begin{equation}
\Delta _F (S, \rho, \tau) = S \left[ F_{2M,2N} - \left( f_{\mathrm{bulk}} + \frac{f_1}{S} + \frac{f_2}{S^2} \right) - \frac{1}{2S} \left( c_b + \frac{1}{2\pi} \ln S \right) \tau^2\right], \label{f_scaling}
\end{equation}
where $\tau$ defined as
\begin{equation}
\tau=t\cdot S^{\frac{1}{2}}, \label{rescaledt}
\end{equation}
is a scaled variable. The scaling function $\Delta _F (S, \rho, \tau)$ as a function of $\tau$ for different system size $S$ with the aspect ratio $\rho=1, 2$, and $4$ is shown in Fig. \ref{fig_7}(a). With the help of the first and second derivatives of the free energy, we obtain the exact expression of the scaling function at criticality for small $\tau$
\begin{equation}
\Delta _F (S, \rho, \tau) = \frac{1}{2} \left[ \frac{c_1}{S} + O \left( \frac{1}{S^2} \right) \right] \tau^2 + O \left( \frac{1}{S^2} \right) + O(\tau^4).
\label{f_scaling_exp}
\end{equation}
For $\rho=1$, we have
\begin{equation}
\Delta _F (S, \rho=1, \tau) = \frac{1}{2} \left[ \frac{0.240428}{S} + O \left( \frac{1}{S^2} \right) \right] \tau^2 + O \left( \frac{1}{S^2} \right) + O(\tau^4),
\label{f_scaling_exp1}
\end{equation}
which describes the behavior of the scaling function at critical region for small $\tau$ in Fig. \ref{fig_7}(a).

We further propose the scaling function of the internal energy $\Delta _U (S, \rho, \tau)$ as
\begin{equation}
\Delta _U (S, \rho, \tau) = S^{\frac{1}{2}} \left[ U_{2M, 2N} - \frac{1}{S^{\frac{1}{2}}} \left( c_b + \frac{1}{2\pi} \ln S \right) \tau \right]. \label{u_scaling_exp}
\end{equation}
The scaling function $\Delta _U (S, \rho, \tau )$ as a function of $\tau$ for different system size $S$ with aspect ratio $\rho=1, 2$, and $4$ is shown in Fig. \ref{fig_7}(b).

Similarly, using the expression of expansion of the specific heat in Eq. (\ref{heatexpansion}), we define the scaling function of the specific heat $\Delta _C (S, \rho, \tau)$
\begin{equation}
\Delta _C (S, \rho, \tau) = C_{2M,2N} - \left( c_b + \frac{1}{2\pi} \ln S \right). \label{c_scaling}
\end{equation}
The scaling function $\Delta _C (S, \rho, \tau)$ as a function of $\tau$ for different system size $S$ with the aspect ratio $\rho=1, 2$, and $4$ is shown in Fig. \ref{fig_7}(c). Note that $\Delta _C (S, \rho, \tau)$ at small $\tau$ can be formulated as
\begin{equation}
\Delta _C (S, \rho, \tau) = \frac{c_{1} (\rho)}{S}+ \left[ \frac{1}{2}g(\rho) -\frac{3}{4\pi}\frac{\ln{S}}{S} + \frac{1}{2}\frac{g_{0}}{S} + O \left( \frac{1}{S^2} \right) \right] \tau^2 + O \left( \frac{1}{S^2} \right) + O(\tau^4), \label{c_scaling_exp1}
\end{equation}
as the case of Ising model \cite{wu2003}, while the leading term of $\tau$ in the scaling function for Ising model is $\tau$. For $\rho=1$,
\begin{equation}
\Delta _C (S, \rho=1, \tau) = \frac{0.240428}{S} - \left[ 0.016061 +\frac{3}{4\pi}\frac{\ln{S}}{S} 
+ O \left( \frac{1}{S^2} \right) \right] \tau^2 + O \left( \frac{1}{S^2} \right) + O(\tau^4). \label{c_scaling_exp2_r1}
\end{equation}
Equations (\ref{f_scaling_exp}), (\ref{u_scaling_exp}), and (\ref{c_scaling_exp1}) and Fig. \ref{fig_7} generally suggest the following relations \cite{wu2003}
\begin{eqnarray}
\Delta _F (S, \rho, \tau) & \simeq &  \frac{1}{2} \Delta _C (S, \rho) \tau^2 + O \left( \frac{1}{S^2} \right) + O(\tau^4), \label{f_and_c_scaling} \\
\Delta _U (S, \rho, \tau) & \simeq & \Delta _C (S, \rho) \tau + O \left( \frac{1}{S^2} \right) + O(\tau^3), \label{u_and_c_scaling} \\
\Delta _C (S, \rho, \tau) & \simeq & \Delta _C (S, \rho) + O \left( \frac{1}{S^2} \right) + O(\tau^2), \label{c_and_c_scaling}
\end{eqnarray}
where $\Delta _C (S, \rho) = \Delta _C (S, \rho, \tau=0)$.

\subsection{Specific heat near the critical point}
\label{spec_heat}

Let us now consider the behavior of the specific heat near the critical point. The specific heat $C_{2M,2N}(t)$ of the dimer model on $2M \times 2N$ checkerboard lattice is defined as the second derivative of the free energy in Eq. (\ref{def_specific_heat}). The pseudocritical point $t_{\mathrm{pseudo}}$ is the value of the temperature at which the specific heat has its maximum for finite $2M \times 2N$ lattice. One can determine this quantity as the point where the derivative of $C_{2M,2N}(t)$ vanishes. The pseudocritical point approaches the critical point $t_c=0$ as $L \to \infty$ in a manner dictated by the shift exponent $\lambda$,
\begin{eqnarray}
|t_{\mathrm{pseudo}}-t_c| \sim L^{-\lambda}. \label{lamb}
\end{eqnarray}
where $L = \sqrt{4MN}$ is the characteristic size of the system. The coincidence of $\lambda$ with $1/\nu$, where $\nu$ is the correlation lengths exponent, is common to most models, but it is not a direct consequence of finite-size scaling and is not always true.

One can see from Eqs.(\ref{stat}), (\ref{twist}) and (\ref{def_specific_heat}) that the partition function $Z_{2M,2N}(t)$ and the specific heat $C_{2M,2N}(t)$ are an even function with respect to its argument $t$
\begin{eqnarray}
C_{2M,2N}(t) = C(0)
+\frac{t^2}{2}C^{(2)}(0)+\frac{t^4}{4!}C^{(4)}(0)+O(t^6).
\label{expansion specific heat}
\end{eqnarray}
Thus the first derivative of $C_{2M,2N}(t)$ vanishes exactly at
\begin{eqnarray}
t_{\mathrm{pseudo}}=0. \label{expansionmu0}
\end{eqnarray}
In Fig. \ref{fig_5}(c) we plot the $t$ dependence of the specific heat for different lattice sizes up to $512 \times 512$. We can see from Fig. \ref{fig_5}(c) that the position of the specific heat peak $t_{\mathrm{pseudo}}$ is equal exactly to zero. Therefore the maximum of the specific heat (the pseudocritical point $t_{\mathrm{pseudo}}$)  always occurs at vanishing reduced temperature for any finite  $2M \times 2N$ lattice and coincides with the critical point $t_c$ at the thermodynamic limit ($t_{\mathrm{pseudo}}=t_c=0$). From Eqs. (\ref{lamb}) and (\ref{expansionmu0}) we find that the shift exponent is infinity $\lambda=\infty$.

\section{Dimer on the infinitely long cylinder}
\label{inf_long}
Conformal invariance of the model in the continuum scaling limit would dictates that at the critical point the asymptotic finite-size scaling behavior of the critical free energy $f_c$ of an infinitely long two-dimensional cylinder of finite circumference ${\mathcal{N}}$ has the form
\begin{equation}
f_c=f_{\mathrm{bulk}}  + \frac{A}{{\mathcal{N}}^2} + \cdots,  \label{freeenergystrip}
\end{equation}
where $f_{\mathrm{bulk}}$ is the bulk free energy and $A$ is a constant. Unlike the bulk free energy the constant $A$ is universal, which may depend on the boundary conditions. In some 2D geometries, the value of $A$ is related to the conformal anomaly number $c$ and the highest conformal weights $\Delta, \bar \Delta$ of the irreducible highest weight representations of two commuting Virasoro algebras. These two dependencies can be combined into a function of the effective central charge
$c_{\mathrm{eff}}=c-12(\Delta+\bar\Delta)$ \cite{Blote,Affleck,Cardy},
\begin{eqnarray}
A = \frac{\pi}{6}c_{\mathrm{eff}} = 2\pi\left(\frac{c}{12}-\Delta-\bar\Delta\right)  \qquad \mbox{on a cylinder.} \label{Aperiod}
\end{eqnarray}
Let us now consider the dimer model on the infinitely long cylinder of width $\mathcal{N} = 2 N$. Considering the logarithm of the partition function given by Eq. (\ref{Zab}) at the critical point ($t=t_{c}=0$), we note that it can be transformed as
\begin{equation}
\ln Z_{\alpha,\beta}(0)= M\sum_{n=0}^{N-1}
\omega_0\!\left(\textstyle{\frac{\pi(n+\alpha)}{N}}\right)+
\sum_{n=0}^{N-1}\ln\left|\,1-e^{-2\big[\,M
\omega_0\left(\frac{\pi(n+\alpha)}{N}\right)-i\pi\beta\,\big]}\right|.
\label{lnZab}
\end{equation}
The second sum here vanishes in the formal limit $M\to\infty$ when the torus turns into infinitely long cylinder of circumference $2N$. Therefore, the first sum gives the logarithm of the partition function on that cylinder. Its asymptotic expansion can be found with the help of the Euler-Maclaurin summation formula
\begin{equation}
M\sum_{n=0}^{N-1}\omega\!\left(\textstyle{\frac{\pi(n+\alpha)}{N}}\right)=
\frac{S}{\pi}\int_{0}^{\pi}\!\!\omega_0(x)~\!{\mathrm{d}}x-\pi\lambda_0\rho\,{\mathrm{B}}_{2}^\alpha-
2\pi\rho\sum_{p=1}^{\infty} \left(\frac{\pi^2\rho}{S}\right)^{p}
\frac{\lambda_{2p}}{(2p)!}\;\frac{{\mathrm{B}}_{2p+2}^\alpha}{2p+2},
\label{EulerMaclaurinTerm}
\end{equation}
where $\int_{0}^{\pi}\!\!\omega_0(x)~\!{\mathrm{d}}x = 2 G$ and ${\mathrm{B}}^{\alpha}_{p}$ are the so-called Bernoulli polynomials. Here we have also used the coefficients $\lambda_{2p}$ of the Taylor expansion of the lattice dispersion relation $\omega_0(k)$ at the critical point given by Eq. (\ref{Spectral}). Thus one can easily write down all the terms of the exact asymptotic expansion for the $F_{\alpha,\beta}= \lim_{M \to \infty}\frac{1}{M}\ln Z_{\alpha,\beta}(0)$
\begin{eqnarray}
F_{\alpha,\beta} &=& \lim_{M \to \infty}\frac{1}{M}\ln Z_{\alpha,\beta}(0)=  \frac{2 G}{\pi}N - 2\sum_{p=0}^\infty
\left(\frac{\pi}{N}\right)^{2p+1}\frac{\lambda_{2p}}{(2p)!} \frac{\mathrm{B}_{2p+2}^{\alpha}}{2p+2}. \label{AsymptoticExpansion1}
\end{eqnarray}

From $F_{\alpha,\beta}$, we can obtain the asymptotic expansion of free energy per bond of an infinitely long cylinder of circumference $\mathcal{N}=2N$
\begin{equation}
f = \lim_{M \to \infty} \frac{1}{4 M N}\ln{Z_{2M,2N}(0)} = \lim_{M \to \infty} \frac{1}{2 M N}\ln {Z_{1/2, 0}(M, N)}=\frac{1}{2 N} F_{1/2,0}(N). \label{free2N}
\end{equation}
From Eq. (\ref{free2N}) using Eq. (\ref{AsymptoticExpansion1}) one can easily obtain that for even ${\mathcal{N}}=2N$ the asymptotic expansion of the free energy is given by
\begin{eqnarray}
f &=& f_{\mathrm{bulk}} - \frac{1}{\pi}\sum_{p=0}^\infty \left(\frac{2 \pi}{\mathcal{N}}\right)^{2p+2}\frac{\lambda_{2p}}{(2p)!} \frac{\mathrm{B}_{2p+2}^{1/2}}{2p+2} \nonumber \\
&=& f_{\mathrm{bulk}} + \frac{\pi}{6}\frac{1}{\mathcal{N}}+\dots \quad ({\mathrm{for}}~ \mathcal{N}=2N), \label{2Nper}
\end{eqnarray}
where $f_{\mathrm{bulk}}$ is given by Eq. (\ref{fbulk}). Thus we can conclude from Eqs. (\ref{freeenergystrip}), (\ref{Aperiod}) and (\ref{2Nper}) that $c_{\mathrm{eff}}=1$. Since the effective central charge $c_{\mathrm{eff}}$ is defined as function of $c$, $\Delta$ and  $\bar \Delta$, one cannot obtain the values of $c$, $\Delta$ and  $\bar \Delta$ without some assumption about one of them. This assumption can be a posteriori justified if the conformal description obtained from it is fully consistent. It is easy to see that there are two consistent values of $c$ that can be used to describe the dimer model, namely, $c = -2$ and $c = 1$. For example, for the dimer model on an infinitely long cylinder of even circumference $\mathcal{N}$, one can obtain from Eqs. (\ref{freeenergystrip}), (\ref{Aperiod}) and (\ref{2Nper}) that the central charge $c$ and the highest conformal weights $\Delta, \bar \Delta$ can take the values $c = 1$ and $\Delta = \bar \Delta = 0$ or $c = -2$ and $\Delta = \bar \Delta = -1/8$. Thus from the finite-size analyses we can see that two conformal field theories with the central charges $c = 1$ and $c = -2$ can be used to describe the dimer model on the checkerboard lattice.

\red{Somewhat surprisingly, these finite-size corrections in the free energy can be consistently interpreted in a conformal scheme based on two conformal descriptions of the dimer model: one with $c = -2$ and the other with $c = 1$. The description of the dimer model in terms of spanning trees on a cylinder \cite{ipph} and on a rectangle \cite{ruelle2007,izmailian2014} supported the $c = - 2$ interpretation, whereas the use of the height function to describe dimer configurations yields $c = 1$ \cite{kenyon,allegra}. Quiet recently Morin-Duchesne, Rasmussen, and Ruelle \cite{ruelle} provided an additional evidence supporting the consistency of a $c = -2$ description of the dimer model on the square lattice. The connection between the dimer model and the critical dense polymer model \cite{rasmussen} adds further support to $c = - 2$. Results from the computation of dimer correlation functions similarly allow for dual representations. Dimer correlations can be naturally accounted for in a $c = - 2$ conformal theory \cite{ruelle2013}, but can also be easily interpreted as $c = 1$ correlators \cite{allegra}. Together, these observations suggest that there exists two conformal descriptions of the dimer model.}

\section{Summary and Discussion}
\label{summary-discussion}

We analyze the partition function of the dimer model on an $2M \times 2N$ checkerboard lattice wrapped on a torus. We have obtained exact asymptotic expansions for the free energy, the internal energy, the specific heat, and the third and fourth derivatives of the free energy of a dimer model on the square lattice wrapped on a torus at the critical point $t=0$. Using exact partition functions and finite-size corrections for the dimer model on finite checkerboard lattice we obtain finite-size scaling functions for the free energy,  the internal energy, and the specific heat of the dimer model.  From a finite size analysis we have found that the shift exponent $\lambda$ is infinity  and the finite-size specific-heat pseudocritical point coincides with the critical point of the thermodynamic limit. This adds to the catalog of anomalous circumstances where the shift exponent is not coincident with the correlation-length critical exponent. We have also considered the limit $N \to \infty$ for which we obtain the expansion of the free energy for the dimer model on the infinitely long cylinder. \red{From a finite-size analysis we have found that the dimer model on the checkerboard lattice can be described by two consistent values of the central charge, namely, $c = -2$ for the construction of a conformal field theory using a mapping of spanning trees and $c = 1$ for the height function description.}

\begin{acknowledgments}
One of us (N.Sh.I.) thanks Laboratory of Statistical and Computational Physics at the Institute of Physics, Academia Sinica, Taiwan,  for hospitality during completion of this work. This work was partially supported by IRSES (Projects No. 612707-DIONICOS)  within 7th European Community Framework Programme (N.Sh.I.) and by a grant from the Science Committee of the Ministry of Science and Education of the Republic of Armenia under Contract No. 15T-1C068 (N.Sh.I.), and by the Ministry of Science and Technology of the Republic of China (Taiwan) under Grant Nos. MOST 103-2112-M-008-008-MY3 and 105-2912-I-008-513 (M.C.W.), and MOST 105-2112-M-001-004 (C.K.H.), and NCTS of Taiwan.
\end{acknowledgments}

\end{document}